# Geometric model of quantum navigation during (anti-)search on a plane

[1,2]A.M. Rostom, [2]V.A. Tomilin, [1,2,3]L.V. Il'ichov

[1]*Novosibirsk State University, 630090, Novosibirsk, Russia*
[2]*Institute of Automation and Electrometry SB RAS, 630090, Novosibirsk, Russia and*
[3]*Institute of Laser Physics SB RAS, 630090, Novosibirsk, Russia*

A model of joint random walk of two agents on an infinite plane is considered. The agents possess no means of mutual classical communication, but have access to quantum entanglement resource which is used according to a pre-arranged protocol. Depending on the details of the protocol, an effective force of attraction or repulsion emerges between the two agents. The emergence of this force from quantum entanglement is interpreted in terms of spherical or hyperbolic geometries for attraction or repulsion, respectively.

**Introduction.** The phenomenon of quantum entanglement – a special type of correlations between parts of a quantum system [1] – defines the essence of the so-called *second quantum revolution*. The bipartite entanglement is broadly used as a potential nonlocal resource [2] for quantum information processing and security protocols [3], quantum imaging, quantum metrology and many more.

The contemporary situation with entanglement is a peculiar one – the abundance of technological applications of entanglement clearly outstrips our level of understanding of its physical essence. This, in general, can be attributed to the controversial status of the quantum state (its ontological or epistemological nature). However, it is an accepted fact that the quantum entanglement remains a real challenge for our physical intuition and can reveal itself in unusual ways.

The present work is dedicated to the study of the role of entanglement in a process of random walk of two agents, *A* and *B*, on an infinite plane and to clarify its underlying geometrical interpretation.

Our entanglement-based random walk protocol proceeds as follows. *A* and *B* agents perform simultaneous steps in random directions. First, they choose random unit vectors $\mathbf{n}_A$ and $\mathbf{n}_B$. Then, they adjust these directions by adding signs $\sigma_A, \sigma_B \in \{\pm 1\}$, so that the final direction is $\sigma_A \mathbf{n}_A$ for *A* and $\sigma_B \mathbf{n}_B$ for *B*. The signs are determined by outcomes of two *local* measurements performed by the agents on their spatially separated and maximally quantum entangled particles (see Fig.1).

The first result of our study is that depending on the scenario the agents agree on before the random walk begins, they experience either effective attraction or effective repulsion. In the former case, the probability of their meeting increases[1], while in the latter case it decreases. These phenomena emerge from entanglement and hence could only be of purely quantum nature.

The second main result is that the effective attraction (repulsion) allows for geometric interpretation by representing them as a result of random walk in a 2D space with positive (negative) curvature.

The influence of entanglement on the choice of strategy of joint actions (including joint search) was first investigated in [4]. Further works [5–7] considered the model with choices of directions $\mathbf{n}_A$ and $\mathbf{n}_B$ from a set in order to achieve superiority over classical random distributions. In these works, the set of specific directions is considered. In contrast, the model considered here includes no restrictions on the chosen directions on a plane. This is the key feature that allows to find the geometric interpretation of effective forces between the wandering agents[2].

To this end, we consider averaged results of a single simultaneous step performed by two agents in case of entanglement-induced correlation and in completely uncorrelated case. Accordingly, we construct an equation that allows to interpret the results in terms of spherical or hyperbolic geometry.

**Joint (anti-)search on a plane.** A standard example of an entangled quantum system, widely used as a first-time demonstration of entanglement and as a basis for many thought experiments, is a pair of spins-1/2 particles *a* and *b* in a singlet state (with zero total spin)

$$|\Psi\rangle = \frac{1}{\sqrt{2}}\Big(|\mathbf{n}\rangle_a \otimes |-\mathbf{n}\rangle_b - |-\mathbf{n}\rangle_a \otimes |\mathbf{n}\rangle_b\Big). \quad (1)$$

Here $|\pm \mathbf{n}\rangle$ are one-particle states with spin projections $\pm 1/2$ along the unit vector $\mathbf{n}$. A notable property of the state $|\Psi\rangle$ is its invariance under the choice of $\mathbf{n}$. Any measurement of particles' spin projection along $\mathbf{n}$ (or any parallel axis) reveals strictly anti-correlated results. If each particle's spin projection is measured along its own direction, $\mathbf{n}_A$ or $\mathbf{n}_B$, then the probabilities of outcomes are

$$p(\sigma_A, \sigma_B | \mathbf{n}_A, \mathbf{n}_B) \doteq |(\langle \sigma_A \mathbf{n}_A | \otimes \langle \sigma_B \mathbf{n}_B |)|\Psi\rangle|^2$$
$$= \frac{1}{4}\Big[1 - \sigma_A \sigma_B (\mathbf{n}_A \cdot \mathbf{n}_B)\Big]. \quad (2)$$

Here $\sigma_A, \sigma_B \in \{\pm 1\}$ are doubled spin projections.

---

[1] By "meeting" we mean *A* and *B* are coming close so that they can detect each other by means of direct observation.

[2] The term "quantum (pseudo)telepathy", borrowed from the quantum game theory and used in [5], is clear, but does not provide a novel physical insight.

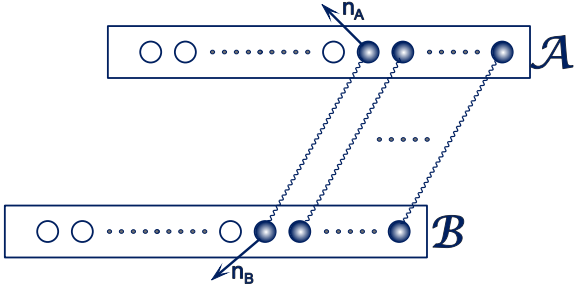

Figure 1. Steps progress of $A$ and $B$ walkers in joint search entanglement-based protocol. The directions $\mathbf{n}_A$ and $\mathbf{n}_B$ for projection measurements are chosen randomly. The wavy line connect particles that form a single entangled pair. The leftmost (bleached) particles describe the already used pairs from the chain.

Suppose that agents $A$ and $B$ posses particles $a$ and $b$, respectively. After choosing random directions $\mathbf{n}_A$ and $\mathbf{n}_B$ on a plane and making measurements on their particles, each agent performs a step of length $l$: $A$ moves along $\sigma_A \mathbf{n}_A$, while $B$ moves along $-\sigma_B \mathbf{n}_B$. Therefore, if the initial distance between them is $r = |\mathbf{r}_A - \mathbf{r}_B|$, then the next one becomes $r' = |\mathbf{r}_A - \mathbf{r}_B + l(\sigma_A \mathbf{n}_A + \sigma_B \mathbf{n}_B)|$. Let us call this joint search algorithm a "plus" protocol. The corresponding mean square distance is

$$\langle r'^2 \rangle_+ = \sum_{\sigma_A, \sigma_B} \int |\mathbf{r}_A - \mathbf{r}_B + l(\sigma_A \mathbf{n}_A + \sigma_B \mathbf{n}_B)|^2 \times p(\sigma_A, \sigma_B | \mathbf{n}_A, \mathbf{n}_B) \frac{d\mathbf{n}_A}{2\pi} \frac{d\mathbf{n}_B}{2\pi} = r^2 + l^2. \quad (3)$$

Here we have performed averaging over uniformly distributed directions $\mathbf{n}_A$ and $\mathbf{n}_B$ and the outcomes of quantum measurements. It is essential that distribution (2) depends on $\mathbf{n}_A$ and $\mathbf{n}_B$.

If $B$ agent makes his steps along $\sigma_B \mathbf{n}_B$, the protocol describes an "anti-search" as we shall shortly see, and will be referred to as "minus" protocol. In this case, the mean square distance becomes

$$\langle r'^2 \rangle_- = \sum_{\sigma_A, \sigma_B} \int |\mathbf{r}_A - \mathbf{r}_B + l(\sigma_A \mathbf{n}_A - \sigma_B \mathbf{n}_B)|^2 \times p(\sigma_A, \sigma_B | \mathbf{n}_A, \mathbf{n}_B) \frac{d\mathbf{n}_A}{2\pi} \frac{d\mathbf{n}_B}{2\pi} = r^2 + 3l^2. \quad (4)$$

Let us compare the results (3) and (4) with a purely classical case, when each agent after choosing a random direction also randomly chooses its sign, e.g. by tossing a coin or following their own free will[3]. Due to the lack of correlations between $\sigma_A$ and $\sigma_B$ the mean square distance, which is defined by the right-hand sides of (3) or (4) with $p(\sigma_A, \sigma_B) = 1/4$ gives

$$\langle r'^2 \rangle_0 = r^2 + 2l^2. \quad (5)$$

---

[3] We assume that $A$ and $B$ have no way to determine some preferred direction $\mathbf{n}_0$ by use of magnetic or gyroscopic devices and then make a biased choice.

Evidently, removing the choice of $\sigma_A$ and $\sigma_B$ entirely would not change the result (5). Note that $\langle r'^2 \rangle_0$ is exactly in between $\langle r'^2 \rangle_-$ and $\langle r'^2 \rangle_+$.

In the "plus", "minus" and classical protocols, the mean square distances have been calculated for a single simultaneous step performed by $A$ and $B$. The number of such steps is increasing at a specific rate with time, and for sufficiently long period of time, it follows from (3), (4) and (5) that the trajectories of $A$ and $B$ will be diverging on average. However, assuming the classical protocol as a standard reference, we can realize that the utilization of particles from a numbered set of singlet pairs (Fig.1) leads to an increase (decrease) in the probability of convergence in the first (second) protocol. This behavior can be understood as effective forces of attraction or repulsion between $A$ and $B$.

Certainly, the word "force" carries a nonliteral meaning. It is thus instructive to formulate a consistent view on the nature of the "attraction" and "repulsion" used here. This, as we shall see below, can be done within a geometric paradigm.

**Geometric model.** We discuss here how $\langle r'^2 \rangle_+$ given by (3) corresponds to $A$ and $B$ walking on a sphere — a two-dimensional surface with positive curvature. This geometry takes the place of the quantum effects and, in principle, the choice of signs $\sigma_A$ and $\sigma_B$, in addition to the choice of random directions, ceases to be necessary, as in the determination of $\langle r'^2 \rangle_0$.

The radius $R$ of the sphere is determined by $r$ and $l$. Consequently, the curvature of space before each next step is determined by the outcome of the previous one. To obtain an equation on $R$, it is convenient to associate with each of the agents, located on the sphere, an individual orthonormal basis: $(\mathbf{e}_{1A}, \mathbf{e}_{2A}, \mathbf{e}_{3A})$, $(\mathbf{e}_{1B}, \mathbf{e}_{2B}, \mathbf{e}_{3B})$. We identify the last elements of the bases with the unit vectors $\mathbf{e}_A$ and $\mathbf{e}_B$, indicating the positions of $A$ and $B$ on the sphere

$$\mathbf{e}_{3A} = \mathbf{e}_A, \quad \mathbf{e}_{3B} = \mathbf{e}_B. \quad (6)$$

It is convenient to make common the first vector of every basis

$$\mathbf{e}_{1A} = \mathbf{e}_{1B} = \mathbf{e}_1 \doteq (\mathbf{e}_B \times \mathbf{e}_A)[1 - (\mathbf{e}_A \cdot \mathbf{e}_B)]^{-1/2}, \quad (7)$$

and the second vectors are obtained automatically

$$\mathbf{e}_{2A} = (\mathbf{e}_A \times \mathbf{e}_1), \quad \mathbf{e}_{2B} = (\mathbf{e}_B \times \mathbf{e}_1). \quad (8)$$

Accordingly, the positions of the agents on the sphere before and after the step become

$$\mathbf{e} = x_1 \mathbf{e}_1 + x_2 \mathbf{e}_2 + x_3 \mathbf{e}_3 = \mathbf{e}_3, \quad (9)$$
$$\mathbf{e}' = x'_1 \mathbf{e}_1 + x'_2 \mathbf{e}_2 + x'_3 \mathbf{e}_3.$$

The subscripts $A$ and $B$ have been omitted here, as will be done further to shorten expressions where appropriate.

Instead of the randomly chosen directions $\mathbf{n}_A$ and $\mathbf{n}_B$ in plane geometry, the movements of A and B on the sphere are specified by axes of rotations $\mathbf{m}_A$ and $\mathbf{m}_B$. Since a straight line segment on a plane is analogous to a great circle arc on a sphere, the axis for each agent lies in the $span\{\mathbf{e}_1, \mathbf{e}_2\}$. The rotation angle is $\theta = l/R$.

The distance $r'$ between A and B after making a single step is determined from the relation

$$\cos(\frac{r'}{R}) = (\mathbf{e}'_A \cdot \mathbf{e}'_B) = \sum_{i,j=1}^{3} x'_{iA} x'_{jB} (\mathbf{e}_{iA} \cdot \mathbf{e}_{jB}). \quad (10)$$

It is convenient to represent triples of coordinates before and after the step in the matrix form

$$\hat{X} \doteq \sum_{i=1}^{3} x_i \hat{\sigma}_i = \hat{\sigma}_3, \quad \hat{X}' \doteq \sum_{i=1}^{3} x'_i \hat{\sigma}_i, \quad (11)$$

where

$$\hat{\sigma}_1 = \begin{pmatrix} 0 & 1 \\ 1 & 0 \end{pmatrix}, \hat{\sigma}_2 = \begin{pmatrix} 0 & -\iota \\ \iota & 0 \end{pmatrix}, \hat{\sigma}_3 = \begin{pmatrix} 1 & 0 \\ 0 & -1 \end{pmatrix},$$

are Pauli matrices. The rotation around the $\mathbf{m}$ axis by the angle $\theta$ is given by the matrix

$$\hat{U}(\mathbf{m}) = \exp\left[-\frac{\iota}{2} \sum_{i=1}^{3} m_i \hat{\sigma}_i \theta\right]. \quad (12)$$

Matrices from (11) are related by the relation

$$\hat{X}' = \hat{U}(\mathbf{m}) \hat{X} \hat{U}(-\mathbf{m}). \quad (13)$$

Directions of the axes are specified by random angles $\varphi_A$ and $\varphi_B$

$$\begin{aligned} \mathbf{m}_A &= \mathbf{e}_1 \cos\varphi_A + \mathbf{e}_{2A} \sin\varphi_A, \\ \mathbf{m}_B &= \mathbf{e}_1 \cos\varphi_B + \mathbf{e}_{2B} \sin\varphi_B. \end{aligned} \quad (14)$$

Determining $\hat{X}'_A$ and $\hat{X}'_B$ from (13) and substituting the result in (10), we obtain

$$\begin{aligned} \cos(\frac{r'}{R}) = &\cos(\frac{r}{R})[\cos\varphi_A \cos\varphi_B \sin^2(\frac{l}{R}) + \cos^2(\frac{l}{R})] \\ &+ \sin(\frac{r}{R})(\cos\varphi_B - \cos\varphi_A)\cos(\frac{l}{R})\sin(\frac{l}{R}) \\ &+ \sin\varphi_A \sin\varphi_B \sin^2(\frac{l}{R}). \end{aligned} \quad (15)$$

Using this expression, we can turn (3) into an equation for $R$

$$R^2 \langle \arccos^2[\cos(\frac{r'}{R})]\rangle = r^2 + l^2. \quad (16)$$

Here, angle brackets stand for averaging over $\varphi_A$ and $\varphi_B$.

It is advisable to construct a model for the "minus" protocol, which leads to repulsion (4), in the space of negative curvature – in hyperbolic geometry on a pseudosphere. The expression corresponding to (15) can be obtained by a formal substitution $R \to \iota R$ (the same trick is used in studies of isotropic cosmological models of the open and closed Universe [8]):

$$\begin{aligned} \cosh(\frac{r'}{R}) = &\cosh(\frac{r}{R})[\cosh^2(\frac{l}{R}) - \cos\varphi_A \cos\varphi_B \sinh^2(\frac{l}{R})] \\ &- \sinh(\frac{r}{R})(\cos\varphi_B - \cos\varphi_A)\cosh(\frac{l}{R})\sinh(\frac{l}{R}) \\ &- \sin\varphi_A \sin\varphi_B \sinh^2(\frac{l}{R}). \end{aligned} \quad (17)$$

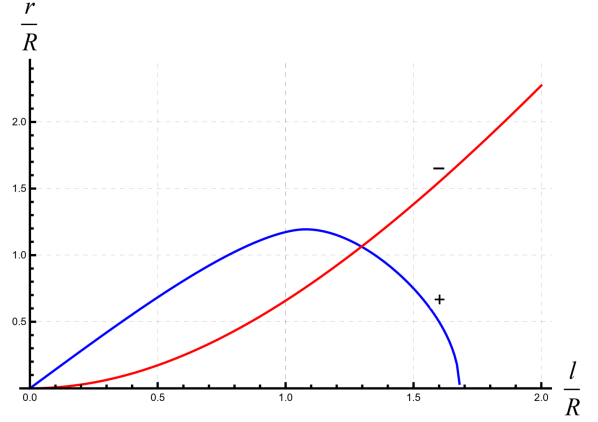

Figure 2. A plot of numerical solutions of (16) and (18) in the form of $r/R$ versus $l/R$. The "+" and "−" curves describe the cases of surfaces having positive and negative curvatures, respectively.

The equation for $R$ now takes the form

$$R^2 \langle \operatorname{arccosh}^2[\cosh(\frac{r'}{R})]\rangle = r^2 + 3l^2. \quad (18)$$

Expression (16), like (18), relates $r/R$ and $l/R$. Numerical solutions expressing the former parameter as a function of the latter are presented in Fig.2. Limited domain for "plus" protocol is natural, because for spherical geometry $r \leq \pi R$. But what is less anticipated is the existence of $\mu$, $0 < \mu < \infty$, such that $r/R \leq \mu l/R$, which indicates that the geometric model for "plus" protocol is only valid for $l/R \geq \mu^{-1}$. This becomes more obvious if we transform the data in Fig.2 to the form of dependence of $R/r$ on $l/r$ (see Fig.3). As $l/r$ approaches $\mu^{-1} \simeq 0.64$ from above, and at finite values of $r$, the sphere radius $R$ increases indefinitely. This corresponds to the transition to plane geometry. The continuation of "+" curve in Fig .2 crosses the horizontal axis at the point $(l/R)_{max}$. From $l/R \leq (l/R)_{max}$, it follows that $R/r \geq (l/R)_{max}^{-1} l/r$, which brings the asymtotics of the corresponding curve in Fig.3. For the curve related to the hyperbolic model, a zero slope at the axis origin is noteworthy.

**Discussion.** Effective forces that appear in (3) and (4) are emerged from the correlation of the pairs of spins in singlet states $|\Psi\rangle$ from (1). In the Werner state

$$\hat{\varrho}(p) = \frac{1-p}{4} \hat{1} \otimes \hat{1} + p |\Psi\rangle\langle\Psi| \quad 0 \leq p \leq 1, \quad (19)$$

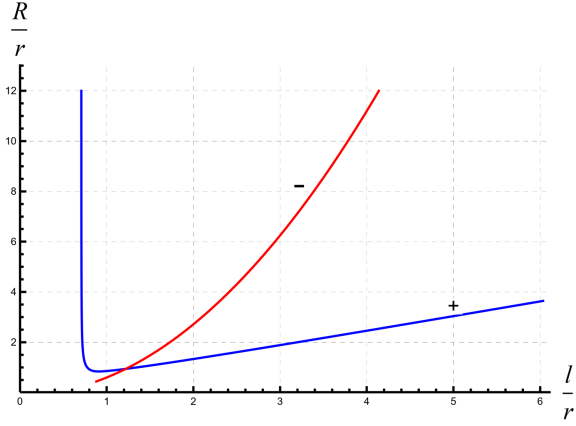

Figure 3. Results from Fig.2 represented as the dependence of $R/r$ on $l/r$. It is clear that the spherical geometry model is inapplicable for $l/r < 0.64$.

which is a mixture of singlet and maximally mixed states of a pair of spin-1/2 particles. The correlations are maximal for $p = 1$ and decrease with $p$ to a minimum of zero. It would be interesting to provide a geometric interpretation for weaker correlations that lead to relations $\langle r'^2 \rangle_\pm = r^2 + w_\pm(p)l^2$. Here $1 < w_+(p) < 2$ and $2 < w_-(p) < 3$ for $0 < p < 1$ (with $w_+(p) + w_-(p) = 4$). In case of spherical geometry, the dependence of $R/r$ on $l/r$ does not experience significant qualitative change upon decrease of $p$. It shifts upwards, towards the region of plane geometry, and to the right, broadening the range of $l/r$ values for which the model does not work. For hyperbolic geometry, the situation is different: for $p < 1$ the slope of $r/R(l/R)$ at the axis origin turns out to be non-zero, i.e. $r/R > \nu l/R$ for some value of $\nu > 0$. Hence, the domain of $R/r$ as a function of $l/r$ has an upper bound: as $l/r$ approaches $\nu^{-1}$, $R/r$ tends to infinity, and for $l/r > \nu^{-1}$ the hyperbolic model as an interpretation of $\langle r'^2 \rangle_- = r^2 + w_-(p)|_{p<1} l^2$ does not work. Maximal entanglement implies the applicability of the hyperbolic model over the entire range of the parameter $l/r$.

It is instructive to illustrate how equations (16) and (18) have a certain similarity to Einstein's equation [8]

$$R_{ij} - \frac{1}{2} R g_{ij} = \frac{8\pi k}{c^4} T_{ij}. \qquad (20)$$

The right-hand side of this equation (the "physical" part) is formed by the energy-momentum tensor $T_{ij}$, while the left-hand side, known as the Einstein tensor $G_{ij}$, has a purely geometric nature. Equations (16) and (18) are much simpler, since they equate scalars, not tensors. The mentioned similarity becomes more evident when investigating energy density in a certain frame. It is given by a set of non-intersecting world lines of observers that span the whole space-time. This frame is related to a field of vectors $\tau^i$ tangent to the world lines of observers (i.e. the field of 4-velocities). The observed energy density is $T_{ij}\tau^i\tau^j$. According to (20), this ("physical") result is up to a constant equal to $G_{ij}\tau^i\tau^j$ – which only depends on the space-time geometry. Similarly, in (16) and (18) the right-hand sides (the "physical" parts) are constrained by the details of quantum protocols, while the left-hand sides depend on the type of effective surface geometries, on which the random walk takes place.

Geometric interpretations of mutual (anti-)search algorithms were based on geometries with maximal symmetry, described by a single parameter $R$. The question about the prospects for more complex anisotropic geometries generated with dedicated geodesics connecting the agents remains open. It is possible that within that framework of such less symmetric models, the above-mentioned upper bound for the range of permissible values of for $l/r$ will be removed.

The curvature of the surface on which the agents move only becomes real for $A$ and $B$ during a joint implementation of the quantum protocol. From the early days of special relativity theory, relating the reality to a certain local observer became a common fact. In our case, the reality – the curvature of space – is relative to a "collective observer" $A \cup B$. This reality originates from the collective observer being involved into a physical process of free choice of the directions ($\mathbf{n}_A$ and $\mathbf{n}_B$), taking into account outcomes ($\sigma_A$ and $\sigma_B$) of quantum measurements conditioned by this choice, performed on the fragments of entangled pairs of particles. Entanglement then plays the role of the resource. It is still unclear how useful and general the introduced concept of the collective observer could be.

In further studies, it is planned to carry out a statistical analysis of the relative movement of agents, implementing both variants of the quantum protocol in comparison with the classical case. It is of interest as well to consider the movement of agents in three-dimensional space or on a real sphere (the planet surface).

*Acknowledgments.* This work was supported by the State order (project AAAA-A21-121021800168-4) at the Institute of Automation and Electrometry SB RAS. Participation of one of the authors (L.V.I.) was supported by RSCF (grant 23-12-00182).

[1] R. Horodecki, P. Horodecki, M. Horodecki, K. Horodecki, Rev. Mod. Phys. **81**, 865 (2009).
[2] N. Brunner *et al.*, Rev. Mod. Phys. **86**, 419 (2014).
[3] C. Portmann and R. Renner, Rev. Mod. Phys. **94**, 025008 (2022).
[4] J. Summhammer, *Quantum Cooperation of Two Insects*, arXiv:quant-ph/0503136 (2005)
[5] C. Brukner, N. Paunkovi, T. Rudolph, and V. Vedral, *Entanglement-assisted Orientation in Space*, arXiv:quant-ph/0509123 (2005).
[6] F.A. Bovino, M. Giardina, K. Svozil, and V. Vedral, *Spatial Orientation using Quantum Telepathy*, arXiv:quant-ph/0603167 (2006).
[7] F.A. Bovino and M. Giardina, Int. J. Quant. Inform., **5**, 43 (2007).
[8] L.D. Landau, E.M. Lifshitz, *Field Theory* (Moscow, Nauka, 1988).